\documentclass[prd,aps,
showpacs,
floatfix,
superscriptaddress,
nofootinbib
]{revtex4}
\usepackage{graphicx}
\usepackage{epsfig}
\usepackage{rotating}
\usepackage{amssymb}
\usepackage{subfigure}
\usepackage{dsfont}
\usepackage{psfrag}
\usepackage{amsmath,euscript,array,mathrsfs}
\usepackage{axodraw}
\usepackage{bbold}
\usepackage{slashed}
\usepackage{color}

\newcommand{\beq}{\begin{equation}}
\newcommand{\eeq}{\end{equation}}
\newcommand{\beqs}{\begin{eqnarray}}
\newcommand{\eeqs}{\end{eqnarray}}

\newcommand{\Tr}{{\rm Tr}}

\def\di{\mbox{d}}

\begin{document}
\title{Calculable mass hierarchies and a light dilaton from gravity duals}

\author{Daniel Elander}
\affiliation{Departament de F\'isica Qu\`antica i Astrof\'isica \& Institut de Ci\`encies del Cosmos (ICC), Universitat de Barcelona, Mart\'i Franqu\`es 1, ES-08028, Barcelona, Spain}

\author{Maurizio Piai}
\affiliation{Department of Physics, College of Science, Swansea University, Singleton Park, Swansea SA2 8PP, UK}

\date{\today}

\begin{abstract}
In the context of gauge/gravity dualities, we calculate the scalar and tensor mass spectrum of the boundary theory defined by a special 8-scalar sigma-model in five dimensions, the background solutions of which include the 1-parameter family dual to the baryonic branch of the Klebanov-Strassler field theory. This provides an example of a strongly-coupled, multi-scale system that yields a parametrically light mass for one of the composite scalar particles: the dilaton. We briefly discuss the implications of these findings towards identifying a satisfactory solution to both the big and little hierarchy problems of the electro-weak theory.
\end{abstract}


\maketitle

\section{Introduction}
\label{Sec:introduction}

Many extensions of the Standard Model (SM) of particle physics are motivated by the
(big) hierarchy problem.
New dynamics and symmetries 
stabilize the electroweak scale, leading to the  expectation 
that new particles should appear just above it.
But such particles have not been detected experimentally
in direct nor indirect searches. The little hierarchy between
the mass of the Higgs and the new particles demands an explanation.

QCD dynamically explains the insensitivity to high-energy scales of the pion decay constant.
New strong dynamics might replicate such success in the electro-weak theory.
However, besides the  calculability  limitations of a strongly-coupled theory,
 the discovery of the Higgs particle~\cite{Higgs} exacerbates the little hierarchy problem in such scenario, as
one would have expected a proliferation of bound states to appear 
above the electroweak scale. 

\begin{figure}[t]
\begin{center}
\includegraphics[height=4cm]{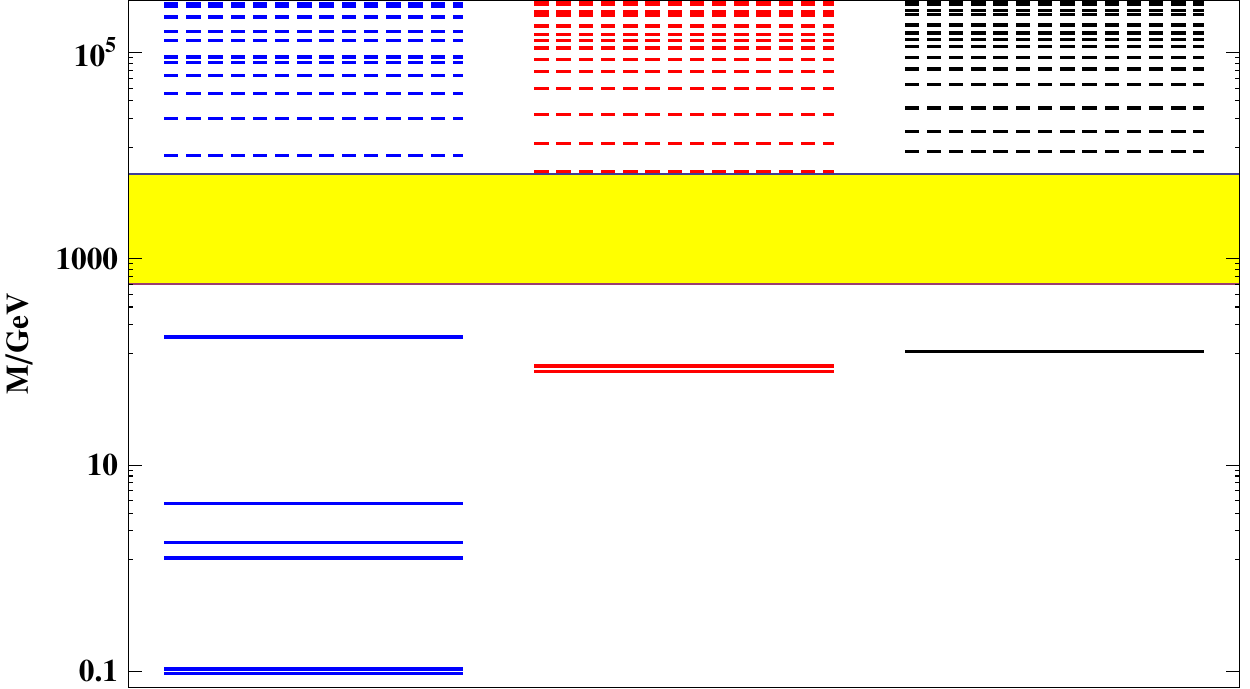}
\caption{ The mass spectrum of SM particles (continuous lines) and
of a generic strongly-coupled new theory (dashed lines)
with new states heavy enough to evade the bounds
 from LHC direct searches  (shaded region)~\cite{ATLAS1,ATLAS2}.
Fermions are rendered in blue, vectors in red and scalars in black.}
\label{Fig:figSM}
\end{center}
\end{figure}

Fig.~\ref{Fig:figSM} provides a pictorial representation of the little hierarchy problem, by showing
the SM mass spectrum,  the current range of bounds from direct searches for 
exotica from the ATLAS collaboration, and the mass spectrum of a generic, 
hypothetic strongly-coupled new  theory that evades  them.
Current bounds range from $570$ GeV (Higgs triplet)~\cite{ATLAS1}
to $6.58$ TeV (Kaluza-Klein graviton)~\cite{ATLAS2}, and (coarse-grained over the details) 
 this range  represents  the LHC reach.
The spectrum  is model dependent,
but consists of infinitely many  bound states of all spins.
If the Higgs scalar and the new physics have a common strong-coupling origin, 
and if the lack of evidence for new physics is confirmed, 
the anomalous suppression of the mass of the Higgs particle must also arise dynamically.

To make strongly-coupled models viable, it is imperative
 to find an example of a strongly-coupled, four-dimensional theory, 
no matter what the microscopic origin,
that exhibits  one scalar state parametrically lighter than
the plethora of bound states.
This possibility arose long ago within walking technicolor~\cite{WTC},
and has been discussed at length in many contexts 
since~\cite{dilatonQFT,dilatonEFT,dilatongravity,dilatonbottom,dilatonlattice}, including 
 Randall-Sundrum models  stabilized \`a la Goldberger-Wise~\cite{RS,GW,GW2}, suggesting that
 the gravity dual of a theory with a moduli space might provide a concrete realisation of a system in which enhanced condensates and hierarchies of scales emerge.
The underlying dynamics  is approximately scale invariant, 
and condensates induce  spontaneous symmetry breaking, 
yielding a light scalar particle in the spectrum: the dilaton.

In this paper, we provide a calculable example of a strongly-coupled theory  
that realizes this scenario, although it does not implement electro-weak symmetry breaking.
Calculability is provided by the regular background in dual
 (super-)gravity~\cite{AdSCFT,MAGOO}: the baryonic branch of the 
Klebanov-Strassler (KS) system~\cite{KS,BGMPZ}. We compute the spectrum via the gauge-invariant fluctuations of the 
background in its 5-dimensional sigma-model description~\cite{BHM1,BHM2,E,EP}.
We refer to~\cite{long} for technical details.
We report the results, discuss their origin, potential applications,
and limitations.

\section{The baryonic branch of KS: field theory}
\label{Sec:KS}

\begin{table}
\begin{center}
\begin{tabular}{|c|c|c|c|c|c|c|}
\hline\hline
{\rm ~~~~~~} &$SU(M)$  &  $SU(M+N)$ & $SU(2)_A$ & $SU(2)_B$ & $U(1)_B$ & $U(1)_R$\cr
\hline
$A_{\alpha}$ & $M$ & $\overline{M+N}$ &$2$ & $1$ &$+1$ & $+1/2$\cr
$B_{\alpha}$ & $\overline{M}$ & $M+N$ &$1$ & $2$ &$-1$ & $+1/2$\cr
\hline\hline
\end{tabular}
\end{center}
\caption{The field content, in terms of chiral superfields, and its  classical symmetries~\cite{KS}.  $SU(M)\times SU(M+N)$ is the gauge group
($M=k N$). An additional $Z_2$ symmetry exchanges $A\leftrightarrow B$ and conjugates the gauge fields.}
\label{Fig:fields}
\end{table}

The four-dimensional ${\cal N}=1$ supersymmetric theory is discussed for example in~\cite{S,MM,GHK,DKS,dimension}.
It has gauge group  $SU(M)\times SU(M+N)$, with $M=k N$ (for $k$ integer), 
and  bifundamental matter fields (see Table~\ref{Fig:fields}). 

The superpotential contains the nearly-marginal:\footnote{The $SU(M)\times SU(M)$ theory is a CFT~\cite{KW}, 
as for $N=0$ there is no anomalous breaking of $U(1)_R\rightarrow Z_{2N}$. The CFT can be obtained as the low-energy IR fixed point reached 
by a mass deformation of the ${\cal N}=2$ supersymmetric gauge theory,
itself obtained by ${Z}_2$-orbifold of the ${\cal N}=4$ theory with gauge group $SU(2M)$.
At the IR fixed point the anomalous dimension is non-perturbative, 
and hence $W$ in Eq.~(\ref{Eq:W}) is not irrelevant.}
\beqs
W&=&h \Tr \left[\frac{}{} A_1B_1 A_2 B_2 - A_1 B_2 A_2 B_1\right]\,.
\label{Eq:W}
\eeqs

The moduli space of the theory contains a baryonic branch.
Following~\cite{DKS,MM}, we illustrate some of its properties with tree-level arguments.
For a fixed choice of $k=q$, the F-term equations are solved by $B_i=0$. We define
\beqs
\Phi_1&=&\left(\begin{array}{cccccc}
\sqrt{q} & 0 & \cdots & 0 & 0 & 0 \cr
0 & \sqrt{q-1} & \cdots & 0 & 0 & 0 \cr
\cdots & \cdots & \cdots & \cdots & \cdots & \cdots \cr
0 & 0 & \cdots & \sqrt{2} & 0 & 0 \cr
0 & 0 & \cdots & 0 & 1 & 0 \end{array} \right)\,,\\
\Phi_2&=&\left(\begin{array}{cccccc}
0 & 1 & \cdots & 0 & 0 & 0 \cr
0 & 0 & \cdots & 0 & 0 & 0 \cr
\cdots & \cdots & \cdots & \cdots & \cdots & \cdots \cr
0 & 0 & \cdots & 0 & \sqrt{q-1} & 0 \cr
0 & 0 & \cdots & 0 & 0 & \sqrt{q} \end{array} \right)\,,
\eeqs
where each element represents a block  proportional to the identity matrix $\mathbb{I}_N$.
The $D$-term equations are
\beqs
0&=&-g_q\sum_i\Tr_{q+1}\left[A^{\dagger}_i T^A_q A_i+B_iT^A_q B^{\dagger}_i\right]\,,\\
0&=&-g_{q+1}\sum_i\Tr_{q}\left[A^{}_i T^A_{q+1} A_i^{\dagger}+B_i^{\dagger}T^A_{q+1} B^{}_i\right]\,,
\eeqs
where the labels $q$ and $q+1$ refer to 
the $SU(qN)$ and $SU((q+1)N)$ groups, respectively, $g_j$ are the gauge couplings, and $T^A_j$ the generators.
Taking $A_i=c \ \Phi_i$:
\beqs
\Tr\left[T^A_q\left(A_1A_1^{\dagger}+A_2A_2^{\dagger}\right)\right]=(q+1)|c|^2\Tr \, T^A_q=0\,.
\eeqs
Another classical branch has $A_i\leftrightarrow B_i$.
The operator
\beqs
{\cal U}&=&\frac{1}{q(q+1)N}\Tr\left[\frac{}{}A_i^{\dagger}A_i-B_iB_i^{\dagger}\right]\,,
\eeqs
normalized so that for $A_i=c \Phi_i$ and $B_i=0$ one has ${\cal U}=|c|^2$,
is the order parameter of $Z_2$ symmetry breaking, and of the Higgsing $SU(qN)\times SU((q+1)N)\rightarrow SU(N)$.

The matrices $\Phi_i$ obey the $SU(2)$ algebra~\cite{MM}, and indeed
 the perturbative calculation of the spectrum of gauge bosons yields $M^2=g^2|c|^2\lambda_{\ell,\pm}$ (for $g_q = g_{q+1} \equiv g$) where
\beqs
\lambda_{\ell,\pm}&=&q+\frac{1}{2}\pm\sqrt{\left(q+\frac{1}{2}\right)^2-\ell(\ell+1)}\,,
\eeqs
where $\ell=0\,,\,1\,,\,\cdots\,,\,q-1$~\cite{MM}, and where the eigenvalues have multiplicity $(2\ell+1)N^2$ for $\ell\neq 0$
and $N^2-1$ for $\ell=0$.
In addition there are $(2q+1)N^2$ states with mass $M^2=g^2|c|^2q$.
The  $N^2-1$ massless vectors represent the unbroken gauge group $SU(N)$. 
The unbroken $SU(N)$ theory with adjoint matter field content can be obtained by twisted compactification 
on a 2-sphere of the CVMN six dimensional field theory~\cite{CVMN}, the degeneracies exactly match, 
while the numerical values of the masses agree for $\ell\ll q$~\cite{AD}.
This theory has a dynamical (confinement) scale $\Lambda$.

The perturbative approach provides a lot of insight~\cite{MM}, but leaves many open questions.
\begin{itemize}
\item The two gauge couplings and the coupling $h$ are not independent, but non-perturbatively related~\cite{S}. 
\item The couplings run, because of the presence of the anomaly. 
The RG flow is best described in terms of a cascade of Seiberg dualities that 
progressively reduce the  group as in 
\begin{widetext}
\beqs
SU(k N)\times SU((k+1)N)\rightarrow SU(k N)\times SU((k-1)N)\rightarrow SU((k-2) N)\times SU((k-1)N)\rightarrow\cdots\,.
\eeqs
\end{widetext}
\item The cascade stops at $k=q$ because of the Higgsing due to ${\cal U}$,
but the constant $c$ should be determined non-perturbatively.
\item Supersymmetry allows  to infer that 
the gaugino condensate forms, breaking $Z_{2N}\rightarrow Z_2$ at scale $\Lambda$,
and to classify the quantum moduli space~\cite{DKS}, but not to
calculate  the  spectrum of bound states. 
\item The K\"ahler part of the supersymmetric action is not protected by non-renormalization theorems,
hence the whole spectrum requires  non-perturbative treatment.
\item The constant $c$ should be linked with $q$. We assume  in the following 
that the position along the (quantum) baryonic branch be characterized by
a non-perturbatively defined $\alpha$, such that $\alpha\rightarrow -\infty$ corresponds to ${\cal U}=0$,
and $\alpha\rightarrow +\infty$ to $q\rightarrow +\infty$.
\item There are two dynamical scales: $\Lambda$ is the scale of explicit symmetry breaking
given by dimensional transmutation of the scale anomaly (beta functions), but ${\cal U}$ is unconstrained, defines a scale that can be taken to be larger than $\Lambda$, and breaks scale invariance spontaneously,
suggesting the presence of  a dilaton in the spectrum, if the latter effect is larger than the former.
\end{itemize}

For all of these reasons, we need a non-perturbative description of the theory at strong coupling,
which is provided (at large $N$) by the known gravity dual~\cite{BGMPZ}.

\section{The baryonic branch of KS: gravity}
\label{Sec:gravity}

The baryonic branch is described in gravity by a  family of type-IIB supergravity backgrounds~\cite{BGMPZ}
(see also~\cite{GMNP,CNP,EGNP})
within the PT ansatz~\cite{PT}, characterized by a compact five-dimensional manifold with the symmetries 
of $T^{1,1}$~\cite{CO}, and a non-compact five-dimensional space with metric
ansatz:
\beqs
\di s^2_5 &=& e^{2A(r)}\di s^2_{1,3}+\di r^2\,.
\label{Eq:metric}
\eeqs
The space can be foliated along $r$ in Minkowski slices, related by a conformal factor 
$e^{2A}$ dependent only on $r$, so that
the radial direction is interpreted in field-theory terms as a renormalization scale.

The general problem of finding solutions and studying their fluctuations can be conveniently formulated in 
terms of a truncation to a five-dimensional sigma-model with $8$ scalars $\Phi^a$ coupled to gravity, 
and Lagrangian
\beqs
{\cal L}&=&
\frac{R}{4}-\frac{1}{2}G_{ab}g^{MN}\partial_{M}\Phi^a\partial_{N}\Phi^b-V(\Phi^a)\,,
\eeqs
where $R$ is the five-dimensional Ricci scalar, $G_{ab}$ is the sigma-model metric, $g_{MN}$ is the metric,
and $V(\Phi^a)$ is the potential. The detailed form of potential and kinetic terms can be found elsewhere~\cite{BHM1}.
The full lift to $10$ dimensions is known~\cite{PT}.\footnote{The five-dimensional system we study is obtained by imposing a set of constraints on the consistent truncation on $T^{1,1}$ in~\cite{CFBGGHO}. In particular, the reduction to eight scalars plus gravity is obtained by imposing a non-linear constraint that is in general not integrable. We treat the resulting sigma-model as a five-dimensional system on its own, and study its fluctuations. To study the full lift to type-IIB supergravity one would need to extend the five-dimensional sigma model along the lines of~\cite{CFBGGHO}, to include additional scalar fields as well as vectors.}

The background solution can be found by using Eq.~(\ref{Eq:metric}) and the assumption that the scalars have non-trivial profiles
$\Phi^a=\bar{\Phi}^a(r)$, and looking for non-singular  (in $10$ dimensions) solutions of the system of coupled differential equations~\cite{BGMPZ}.

The baryonic branch solutions are characterized by  two scales $r_0$ and $\bar{r}$,
the separation of which is controlled by the integration constant $\alpha$~\cite{long}.
When $r>\bar{r}$, the background is  approximated by the KS solution, and captures the cascade of Seiberg dualities for $k>q$.
For $r<\bar{r}$  the background approaches the CVMN solution~\cite{CVMN}: 
the quiver gauge theory is higgsed to the $SU(N)$ one.
As $r \rightarrow r_0$, the space ends: the theory confines and the gauginos condense.
The gravity background captures all the non-perturbative features expected from field theory.

In computing the spectrum, we restrict our attention to tensor and scalar modes
by fluctuating  the five-dimensional sigma-model around the background solutions.
We employ the gauge-invariant formalism of~\cite{BHM1,BHM2,E}. The
tensor $\mathfrak{e}^{\mu}_{\,\,\,\nu}$ is the transverse and traceless part of the fluctuations of the four-dimensional metric.
The gauge-invariant scalars $\mathfrak{a}^a$ are written in terms of the fluctuations $\varphi^a$ of the bulk scalars
and $h$ of the trace of the four-dimensional part of the metric:
\beqs
\mathfrak{a}^a&=&\varphi^a-\frac{\partial_{r}\bar{\Phi}^a}{6\partial_r A}h\,.
\eeqs
The tensors obey the linearized differential equations 
\beqs
0&=&\left[\frac{}{}\partial_r^2+4\partial_rA\partial_r+e^{-2A}m^2\right]\mathfrak{e}^{\mu}_{\,\,\,\nu}\,,
\eeqs
where $m$ is the four-dimensional mass, and for the scalars
\begin{widetext}
\beqs
\label{Eq:diffeq}
	0&=&\Big[ {\cal D}_r^2 + 4 \partial_{r}A {\cal D}_r + e^{-2A} m^2 \Big] \mathfrak{a}^a \\ \nonumber
	&& - \Big[ V^a_{\ |c} - \mathcal{R}^a_{\ bcd} \partial_{r}\bar\Phi^b \partial_{r}\bar\Phi^d + \frac{4 (\partial_{r}\bar\Phi^a V^b + V^a 
	\partial_{r}\bar\Phi^b) G_{bc}}{3 \partial_{r} A} + \frac{16 V \partial_{r}\bar\Phi^a \partial_{r}\bar\Phi^b G_{bc}}{9 (\partial_{r}A)^2} \Big] \mathfrak{a}^c\,.
\eeqs
\end{widetext}
In order to interpret the eigenstates as states in the dual theory, we impose the boundary conditions:
\beqs
\left.\frac{}{}\partial_r\mathfrak{e}^{\mu}_{\,\,\,\nu}\right|_{r=r_i}&=&0\,,
\eeqs
and~\cite{EP}
\beqs
\label{Eq:BCb}
 \frac{2  e^{2A}\partial_{r}\bar \Phi^a }{3m^2 \partial_{r}A}
	\left[ \partial_{r}\bar \Phi^b{\cal D}_r -\frac{4 V \partial_{r}\bar \Phi^b}{3 \partial_rA} - V^b \right] \mathfrak a_b +\mathfrak a^a\Big|_{r_i} = 0, \ \
\eeqs
where $r_i=r_I,r_U$ are cutoffs, acting as regulators.

\begin{figure}[t]
\begin{center}
\includegraphics[height=5.0cm]{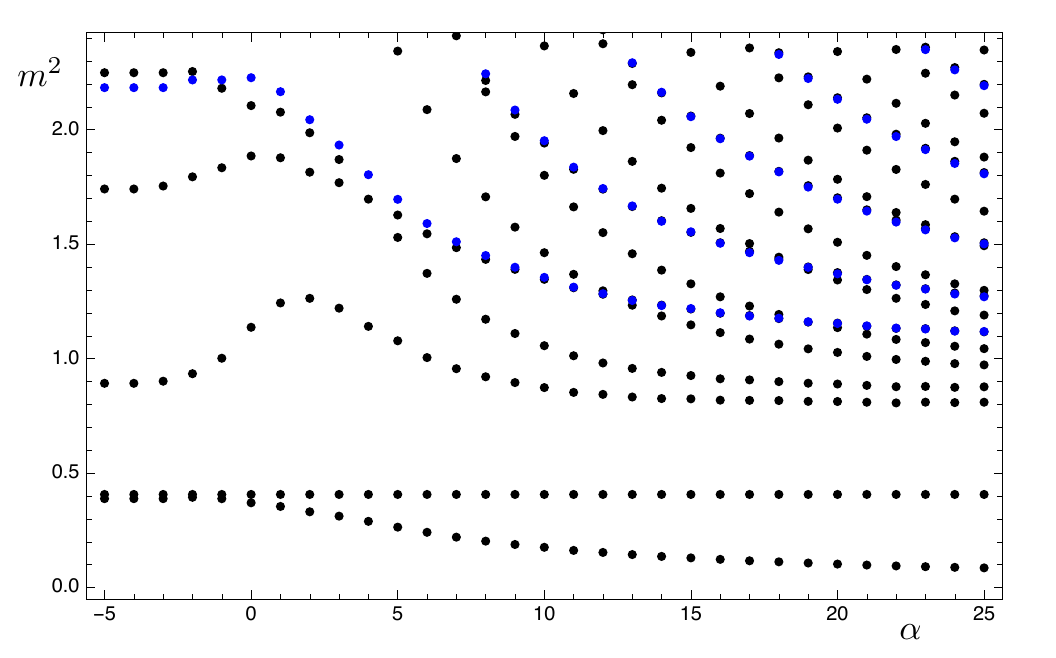}
\caption{The mass spectrum $m^2$ of scalar (black) and tensor (blue) states.}
\label{Fig:spectrum}
\end{center}
\end{figure}

The two regulators have no physical meaning: they are needed in the numerical calculation for practical reasons,
but the physical results are obtained by extrapolating to $r_I\rightarrow r_0$ and $r_U\rightarrow +\infty$~\cite{long}.
Here we report only the final physical results, contained in
Fig.~\ref{Fig:spectrum}, where we show the spectrum of scalar and tensor modes as a function of the parameter $\alpha$ characterizing 
the position along the baryonic branch. We are interested only in ratios of masses, to facilitate the comparisons
 we chose the normalization so that for all values of $\alpha$ the next-to-lightest state 
agrees with the lightest state of the CVMN spectrum~\cite{BHM2}.

The main results that emerge are the following.
\begin{itemize}
\item Scalar and tensor spectra agree with the KS case for small $\alpha$~\cite{BHM1,spectrumKS},
and show evidence of deconstruction at the non-perturbative
level: at  large  $\alpha$, part of the spectrum consists of
a dense sequence of states, above the continuum thresholds of the CVMN case
 ($\alpha\rightarrow +\infty$)~\cite{BHM1,BHM2}.
\item The spectrum of scalars contains one state the mass of which is suppressed going to large $\alpha$. In the limit $\alpha\rightarrow +\infty$, we expect this state to become massless and decouple, as suggested by Fig.~\ref{Fig:spectrum} and by the fact that this state is absent in the spectrum of the CVMN background, in which case it does not correspond to a normalizable, massive mode~\cite{long}.
\end{itemize}
The latter is the main element of novelty of this paper.

\section{Discussion}
\label{Sec: discussion}

All the qualitative expectations emerging from the study of the field theory are confirmed quantitatively
by the gravity dual, including the 
fact that the spectrum of glueballs along the baryonic branch deconstructs a compact manifold, 
and interpolates between the known spectra of 
the KS and CVMN backgrounds.

In addition, we find a new result not accessible using field theory methods: the spectrum of scalars contains one parametrically light 
state,  the mass of which is suppressed moving far from the origin of the baryonic branch. We interpret it as a dilaton,
on the basis of the fact that all the solutions considered here are dual to the same field theory, with scale $\Lambda$
controlled by the explicit breaking of scale invariance, but differ by the tunable choice of the
vacuum value of the operator ${\cal U}$. 

The dynamical scale $\Lambda$ 
 is controlled by explicit symmetry breaking (the scale anomaly), 
and hence is natural.
There emerges in addition a tunable hierarchy between the mass of one
isolated scalar (the dilaton) and the typical scale of the other states.
This is hence an example of a strongly-coupled theory that naturally provides both a big and a little hierarchy of scales,
thanks to the role of scale invariance and to the hierarchy between the scales of its spontaneous and explicit breaking.

A complete and rigorous understanding of the field theory requires extending the formalism for treating the fluctuations to adapt it to a more general truncation including vectors in the sigma model~\cite{CFBGGHO}.
We leave this task for future work. Furthermore, the theory is supersymmetric, and the gravity calculation captures only  the large-$N$ limit. 
The latter provides a technical advantage, as it allows to perform a conceptually simple
(although numerically challenging) calculation. 
The dilaton is parametrically light because the theory admits a classical moduli space
along a flat direction that is lifted only by controllable quantum effects (the running of the couplings), 
with the quantum moduli space still non-trivial~\cite{DKS}, and hence there are condensates
that are parametrically larger than the scale introduced by the explicit breaking of scale invariance due to the anomaly. Whether such phenomena arise in non-supersymmetric theories is an open problem.

The results of this paper support the expectation that in a strongly-coupled theory in which condensates are enhanced, the mass spectrum of bound states (including the lightest scalar) would reproduce the qualitative features in Fig.~\ref{Fig:figSM}. A phenomenologically viable solution of the hierarchy problem(s) of the electro-weak theory would require to implement electro-weak symmetry and its breaking. {One possible way to achieve this, within the specific context of the model studied here, might follow the lines of the Sakai-Sugimoto model~\cite{SS},
 by embedding in the background extended objects~\cite{D-TC,A}
that implement the $SU(2)\times SU(2)$ global symmetry of the SM Higgs sector,
and its breaking. The gauging of $SU(2)_L\times U(1)_Y$ must be reinstated via holographic renormalization, by including boundary-localised terms that cancel the divergence of the gauge boson wave function, and hence retain a finite gauge coupling in the limit in which the UV cutoff is taken to infinity.
While finding the embedding is technically challenging~\cite{DKS2},
given the great potential of this or alternative model-building approaches, 
we think this line of reasoning deserves further future study.

\vspace{1.0cm}
\begin{acknowledgments}

We would like to thank D.~Mateos and C.~Nunez for useful discussions, and A.~Faedo for important comments regarding the non-linear constraint of the sigma-model. DE is supported by the ERC Starting Grant HoloLHC-306605 and by the grant MDM-2014-0369 of ICCUB.
The work of  MP is supported in part  by the STFC grant ST/L000369/1.

\end{acknowledgments}



\end{document}